\documentclass[usenatbib]{basi}
\usepackage[T1]{fontenc}
\usepackage[british]{babel}
\usepackage[varg]{txfonts}
\usepackage{hyperref}
\begin{document}

\title[Gaia FGK Benchmark stars]{Gaia FGK benchmark stars: a bridge between spectroscopic surveys}

\author[P. Jofr\'e]{Paula Jofr\'e$^{1,2}$\thanks{email: \texttt{pjofre@ast.cam.ac.uk}}, 
   Ulrike Heiter$^3$ and Sven Buder$^4$\\
    $^1$ Institute of Astronomy, University of Cambridge, Madingley Road, Cambridge CB3 0HA, UK\\
    $^2$ N\'ucleo de Astronom\'ia, Universidad Diego Portales, Ej\'ercito 441, Santiago, Chile.\\
    $^3$ Department of Physics and Astronomy,  Uppsala University, Box 516, 75120 Uppsala, Sweden\\
    $^4$ Max-Planck Institute for Astronomy, 69117, Heidelberg, Germany}

\pubyear{2012}
\volume{00}
\pagerange{\pageref{firstpage}--\pageref{lastpage}}

\date{Received --- ; accepted ---}

\maketitle

\label{firstpage}

\begin{abstract}
The {\it Gaia} benchmark stars (GBS) are very bright stars of different late spectral types, luminosities and metallicities. They are well-known in the Galactic archaeology community because they are widely used to calibrate and validate the automatic pipelines delivering parameters of on-going and future spectroscopic surveys. 
The sample provides us with consistent fundamental parameters as well as a library of high resolution and high signal-to-noise spectra. This allows the community  to study details of high resolution spectroscopy and to compare results between different survey pipelines, putting the GBS at the heart of this community.  
Here we discuss some results arising from using the GBS as main data source for spectral analyses.  

%The {\it Gaia} benchmark stars are few very bright and well-known stars that are used to calibrate and validate stellar parameters in a large fraction of on-going spectroscopic surveys. They are also used to understand detailed aspects of spectral analysis. Examples are shown in this article.\\[6pt]

\end{abstract}


%------------------------------------------------------------------------------%
\section{Introduction}\label{sec:intro}

%One goal of the Galactic archaeology community is  to complement the photometry, astrometry, spectroscopy and seismology of large samples of stars to have distances, proper motions,  radial velocities, fundamental stellar parameters and individual chemical abundances. This information is then used to study the spacial, dynamical and chemical distributions of stars of different ages and at different locations, which are needed to unravel the structure and evolution of the Galaxy. 

One goal of the Galactic archaeology community is to unravel the structure and evolution of the Galaxy. This is approached by studies of the spacial, dynamical and chemical distributions of stars of different ages at different Galactic directions, based on the photometry, astrometry, spectroscopy and seismology of large samples of stars complemented by distances, proper motions, radial velocities, fundamental stellar parameters and individual chemical abundances.

In the era of ultra-high precision stellar astrophysics opened up by the \emph{CoRoT}, \emph{Kepler}, and \emph{Gaia} missions, there are many spectroscopic surveys available to the community today, notable examples are RAVE, SDSS, LAMOST, Gaia-ESO, APOGEE and GALAH. In the future, high-resolution spectroscopic surveys like WEAVE and 4MOST dedicated to follow up Gaia sources will provide even larger datasets.   These different surveys are designed to observe different kind of stars, and therefore their spectra are of different nature. Some cover the optical wavelength range, others cover infrared. Some target the nearby solar neighbourhood, others the faint distant halo. The resolution,  signal-to-noise, and wavelength coverage, also varies from one survey data to another. This naturally leads to different and independent  automatic pipelines developed for each of these surveys. Some of these pipelines estimate parameters with spectrum syntheses, others use equivalent width measurements. 

The ideal way to compare the results of these pipelines, which gives a handle of the systematic uncertainties of these results, is to test the pipelines against sets of references stars {\it in common}. The GBS is one example of a very suitable and successful set of common stars. Other examples are stars in well-known clusters like M67, and in fields with asteroseismic observations \citep[see e.g.][]{pancino}%In that sense, the GBS and other reference stars constitute eh link between the ``broad sweeper" and the ``ultimate refiner", as suggested by E. Griffin in this conference.  

In the IWSSL 2017 workshop, several issues of stellar spectroscopy that have arisen from the analysis of GBS were discussed  in the context of large spectroscopic surveys, Gaia and the future of Galactic archaeology.  We start with a short description of the GBS, then we summarise two of our recent activities and we finalise with future prospects. 

\section{The FGK Gaia Benchmark stars}

The GBS  are among the main calibrators of the Gaia-ESO Survey and are a central source for validation of many recent independent spectroscopic analyses. Documentation of the work can be found in the series of six A\&A articles published between 2014 and 2017. In these articles one finds their selection criteria \citep{paperI}, a public spectral library \citep{paperII}, their spectral analyses for metallicity \citep{paperIII} and $\alpha$ and iron-peak abundances \citep{paperIV}.  Five new metal-poor candidates were included to  the GBS list in \cite{paperV} and recently a study of the systematic uncertainties in abundance determinations was presented in \cite{paperVI}. 

The final sample with its recommended parameters is available in the website of our  library ({\url{https://www.blancocuaresma.com/s/benchmarkstars}}).

\section{Validation/calibration fields for spectroscopic surveys}

Recently, having large numbers of common targets between different spectroscopic surveys became one of the first priorities as this allow to perform direct comparisons of the different pipelines. This is necessary to improve the performance of the pipelines and to develop a strategy to put the parameters of the different surveys into the same scale. Furthermore, the larger the sample of stars, the better  for developing data-driven methods which transfer the information (e.g. stellar parameters and chemical abundances) from one dataset onto another. So far, still a limited number of common stars between surveys is publicly available to the community.

\begin{figure}[t]
\begin{center}
\includegraphics[width = \columnwidth]{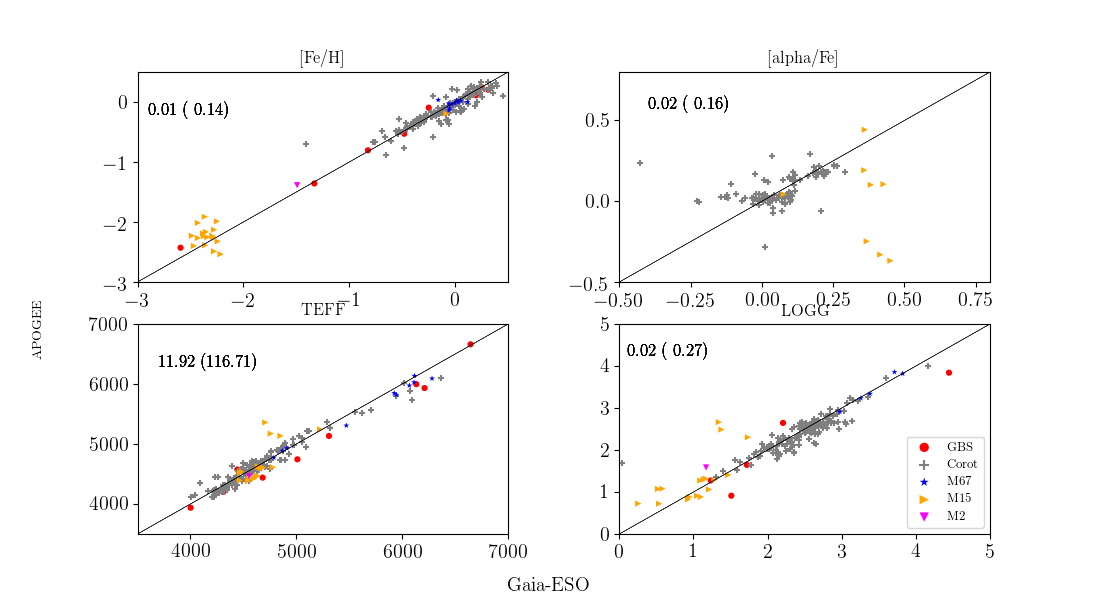}
\caption{Comparison of stellar parameters of common stars observed by APOGEE and GES.}
\label{params}
\end{center}
\end{figure}

In July 2016 a group of specialists involved in survey spectral parametrisation pipeline development agreed to meet in Sexten at a workshop entitled ``Industrial Revolution of Galactic Astronomy" organised by A. Miglio, P. Jofr\'e, L. Casagrande and D. Kawata. During a week the group compared and discussed the parameters and abundances obtained by different survey pipelines. There are about 200 stars in common between Gaia-ESO Data Release 4 and APOGEE Data Release 13  whose parameters can be obtained from public databases\footnote{\url{http://www.eso.org/qi/} and \url{http://www.sdss.org/dr13/irspec/} for GES and APOGEE, respectively.}. These common stars correspond to a subset of GBS, cluster members and CoRoT targets. The comparison of these parameters is shown in Fig.~\ref{params}, with each symbol representing a different set of stars. At the left top of each panel the mean difference  is indicated, with its standard deviation in parenthesis. 

Temperature, surface gravity and metallicity show a very good agreement with negligible systematic differences and a scatter that is comparable with typical errors of such parameters. This is encouraging since both pipelines have been developed in completely independent ways, employing a very different strategy and focused in spectral windows with no overlap. The APOGEE pipeline is estimating the best parameters by chi2-optimisation with a pre-computed grid of synthetic spectra in the infrared. \citep[see][]{garciaperez}. In contrast, the GES parameters and abundances are the product of the homogenisation of multiple independent pipelines, which employ a variety of approaches, including excitation and ionisation equilibrium estimation based on equivalent widths as well as  chi2-optimisation of synthetic spectra in the optical. \citep[][Hourihanne et al in prep]{smiljanic}. The alpha abundances, however, do not show a clear correlation like the other parameters, and the reason for this was investigated further.

\begin{figure}[t]
\begin{center}
\includegraphics[scale = 0.30]{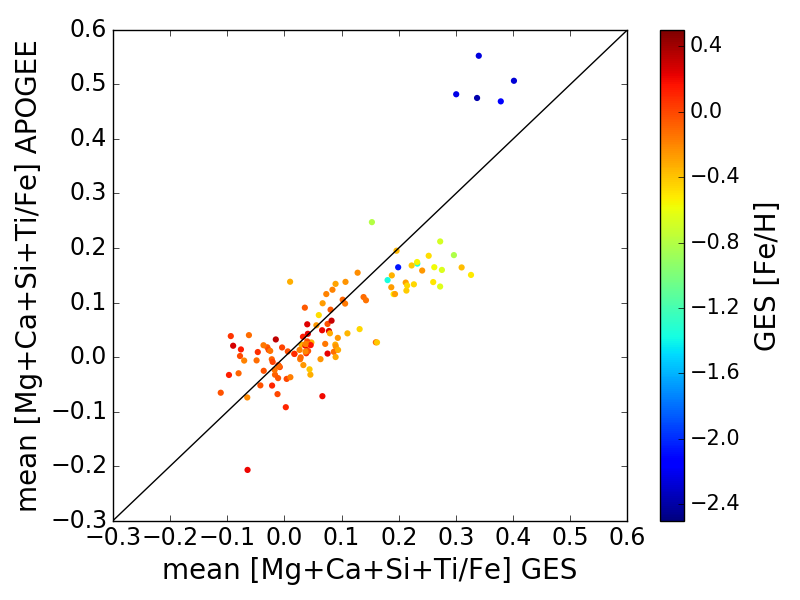}
\includegraphics[scale = 0.30]{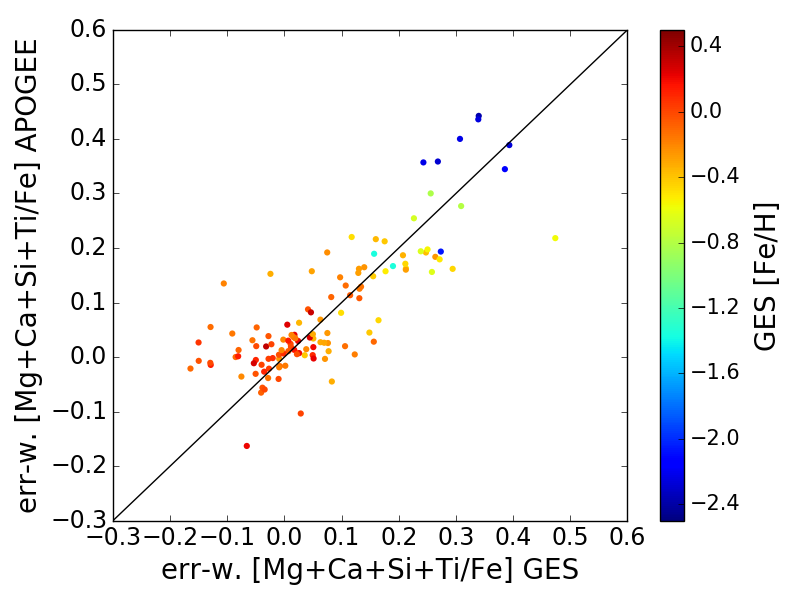}
\caption{Comparison of alpha abundances between AGOGEE and GES.}
\label{alphas}
\end{center}
\end{figure}

It is important to clarify that the [$\alpha/$Fe] values reported by GES and APOGEE correspond to the value considered in the atmosphere model that reproduces the synthetic spectrum that best fits the data. 
The value is however based on the $\alpha-$element features dominating the respective spectral regions of the two surveys. 
%Thus, this value essentially represents a combination of the different $\alpha$ elements and can vary from region to region in the spectrum according to the weak or strong presence of lines of the different $\alpha$ elements.
The APOGEE spectra have for example numerous very strong oxygen features, while the GES spectra is dominated by Mg and Ca features. Since the $\alpha$ elements are produced by slightly different nucleosynthesis processes, it is not surprising that a star might have slightly different abundances of O and Mg, for example. The fact that APOGEE and GES's [$\alpha/$Fe] parameters are not tightly correlated does not mean that one of the datasets have incorrect values, but reflects that the $\alpha$ signatures differ in the different spectral domains.    Figure~\ref{alphas} shows the comparison of individual $\alpha$-elements that have been commonly measured by GES and APOGEE for the same stars as shown in Fig.~\ref{params}. This removes some of the extreme outliers seen in Fig.~\ref{params}. The correlation improves even more when the abundances are weighted according to the precision with which they are measured (right hand panel). 

It remains to be seen which is the best way for spectroscopic surveys to provide an $\alpha$ abundance parameter, such that different stellar populations can be properly identified, as well as different surveys properly scaled. This test shows the importance of making clear of what the $\alpha$ parameter means by each of the surveys.

\section{The art of stellar spectroscopy}

It has been argued that modern astrophysics started with the advent of stellar spectroscopy in the 19th century \citep{Becker:93}\footnote{\url{http://faculty.humanities.uci.edu/bjbecker/huggins/}}.
 A remarkable example of stellar spectroscopy is the creation of the Henry Draper (HD) Catalogue, in which the {\it Harvard Computers} (women hired to perform computations in Harvard) classified by eye more than 200,000 stellar spectra according to the depth of some lines leading to the sequence of spectral types we employ today.    While this classification is still used today, the amount and quality of data has grown massively recently. The level of detail with which we can resolve spectral lines  today with current high-resolution spectrographs is impressive. At the same time, our computers (now machines performing computations) are capable of solving quite complicated equations of stellar structure including 3D modelling of atmospheres, rotation and departures from local thermo-dynamical equilibrium. Thus,  we can achieve a much higher accuracy and much more objective classification of stars than simply by eye.

However, it has become evident that even in modern times there is much room for improvement in the art of interpreting stellar spectra. It has been shown that different input material, namely atmosphere models, atomic and molecular data, and observed spectra --which have different resolutions, wavelength coverage, signal-to-noise ratios (SNR)-- might cause different results in derived stellar parameters, in particular in metallicity and individual abundances. By fixing these variables one might be able to quantify the systematic uncertainties in stellar parameters when different methodologies are based on the same input material.
 
The Gaia-ESO Survey is the first and only project that has attempted, massively and systematically, to employ multiple available techniques on a very large dataset. Its approach has been to combine the results results of about 20 different groups analysing simultaneously several thousands of spectra in about 5 different wavelength regions and resolutions. A final value and a statistical uncertainty for the stellar parameters of each star are provided (see talk of R. Smiljanic in this conference). To the surprise of many,  this uncertainty has turned out to be much larger than expected in some cases.  %In fact, in some cases it is so large that we began to doubt if modern stellar spectroscopy is such a well-established field of research after all! 
 
 Is it only the different choices of spectral lines and the possibly insufficient SNR of the observations which are to blame for the discrepancies in obtained stellar parameters?
  That is one of the questions we try to answer using the GBS, for which we have spectra at exquisite SNR, resolution and wavelength coverage. In February 2016 a workshop in Cambridge was organised to tackle this question and the results were published in \cite{paperVI}. We could  quantify the differences obtained in abundance determination with different methods based on the same spectral lines, atmosphere models, stellar parameters and atomic data. The methods considered used using state-of-the art tools based on synthesis and equivalent widths. Essentially we investigated to what extent the ``default" parameters of each method (continuum placement, model interpolation, continuum opacities, etc) affect the final abundances.   We found that differences in continuum normalisation, even in very high SNR spectra,  caused an impact of up to 0.6~dex in the retrieved abundances, while weak blends or interpolation of model atmospheres had insignificant effects on final abundances.

\section{Future prospects}
The GBS work is continuing to progress towards different directions simultaneously. Below we list some of them. 
\begin{itemize}
\item  {\bf Hunting new and better candidates}. Several interferometric programmes are ongoing to enlarge the sample of stars and to improve the accuracy of measured angular diameters. With Gaia DR2 parallaxes and the new interferometric data,  we will revise our GBS sample and provide a new set of parameters.
\item {\bf Improving spectral library and line list for analysis}. We have included spectra of the entire optical range using all setups of UVES archival data.  This has been distributed to a small group of people to do abundance analysis (below) but will be provided soon in our Library webpage. %We hope to include infrared spectra taken with the Carmenes telescope by D. Montes (Madrid) in the near future. 
\item {\bf Determination of abundances}. We are analysing the new GBS library with the goal to provide reference abundances of light elements (Li, C, N, O, Al, Na) and heavy elements (tbd). For this purpose,  extensive work  is on-going in improving the atomic line list outside the GES range starting from the data available in the VALD database\footnote{\url{http://vald.astro.uu.se}}. Furthermore, significant effort had to be invested to select the best ranges to determine abundances of C,N and O from molecular bands. 
%\item {\bf Cross-calibration of Surveys} Attempts to compare parameters and abundances of the observed GBS together with other CoRot fields and M67 stars between APOGEE, GES and GALAH have were carried out in a workshop in Sexten in July 2016 (see Fig.~\ref{alpha-combo}). Parameters show good agreements but abundances are more scattered. Continuing this work would enable the improvement of parameters delivered by all surveys.  GBS alone are however too few. We are looking for twin stars of GBS (Worley, Jofre) to provide new validation fields of fainter GBS that could be observed by surveys easier and be used for these kind of studies. 
\end{itemize}

The future of Galactic archaeology is moving towards analyses of very large combined datasets provided by future surveys providing with parallaxes, proper motions and radial velocities (Gaia), colours (LSST), stellar parameters and chemical abundances (4MOST and WEAVE)  and masses and ages (K2 and PLATO). At the same time, the extremely high levels of details that we can detect in high-quality spectra of stars in our solar neighbourhood tell us how complex a star like the Sun or Arcturus can be, challenging the finest physical assumptions involved in the theory of stellar structure and evolution. As insisted during this conference by E. Griffin, a proper connection between the ``wide sweeper" and the ``ultimate refiner" is now more important than ever.


\begin{thebibliography}{}
%

\bibitem[Heiter et al.(2015)]{paperI} Heiter, U., Jofr{\'e}, P., Gustafsson, B., et al.\ 2015, A\&A, 582, A49 
\bibitem[Blanco-Cuaresma et al.(2014)]{paperII} Blanco-Cuaresma, S., Soubiran, C., Jofr{\'e}, P., \& Heiter, U.\ 2014, A\&A, 566, A98 
\bibitem[Jofr{\'e} et al.(2014)]{paperIII} Jofr{\'e}, P., Heiter, U., Soubiran, C., et al.\ 2014, A\&A, 564, A133  
\bibitem[Jofr{\'e} et al.(2015)]{paperIV} Jofr{\'e}, P., Heiter, U., Soubiran, C., et al.\ 2015, A\&A, 582, A81 
\bibitem[Hawkins et al.(2016)]{paperV} Hawkins, K., Jofr{\'e}, P., Heiter, U., et al.\ 2016, A\&A, 592, A70 
\bibitem[Jofr\'e et al.(2017)]{paperVI} P. Jofr\'e, U. Heiter, C. C. Worley, S., et al. \ 2017, A\&A, 601, A38 
\bibitem[Pancino et al.(2017)]{pancino} Pancino, E., Lardo, C., Altavilla, G., et al.\ 2017, A\&A, 598, A5 
\bibitem[Garc{\'{\i}}a P{\'e}rez et al.(2016)]{garciaperez} Garc{\'{\i}}a P{\'e}rez, A.~E., Allende Prieto, C., Holtzman, J.~A., et al.\ 2016, AJ, 151, 144 
\bibitem[Smiljanic et al.(2014)]{smiljanic} Smiljanic, R., Korn, A.~J., Bergemann, M., et al.\ 2014, A\&A, 570, A122 
\bibitem[Becker (1993)]{Becker:93} Becker, B.\ 1993, PhD Thesis,  The Johns Hopkins University, Baltimore, MD%, ``Eclecticism, Opportunism, and the Evolution of a New Research Agenda: William and Margaret Huggins and the Origins of Astrophysics"











\end{thebibliography}
\end{document}